# On The Secrecy of the Cognitive Interference Channel with Channel State

Hamid G. Bafghi, Babak Seyfe
*Department of Electrical Engineering, Shahed University, Tehran, Iran.*
*Emails: {ghanizade, seyfe}@shahed.ac.ir*

*Abstract*—In this paper the secrecy problem in the cognitive state-dependent interference channel is considered. In this scenario we have a primary and a cognitive transmitter-receiver pairs. The cognitive transmitter has the message of the primary sender as side information. In addition, the state of the channel is known at the cognitive encoder. So, the cognitive encoder uses this side information to cooperate with the primary transmitter and sends its individual message confidentially. An achievable rate region and an outer bound for the rate region in this channel are derived. The results are extended to the previous works as special cases.

I. INTRODUCTION

Contemporaneous of arising the interference channels as a basic model in studying the communication literature, using the channel state in communication channel models was introduced by Shannon in his landmark paper [1]. He assumed the channel side information at the transmitter (CSIT). Gelfand and Pinsker in their essential work [2] proved that the capacity of the discrete memoryless channel with non-causal CSIT is given by $C = \max_{p(u,x|v)}[I(U;Y) - I(U;V)]$, where the maximum is taken over all input distributions $p(u,x|v)$ with a finite alphabet auxiliary random variable $U$.

Costa in his famous paper, named *writing on dirty paper*, extended Gelfand and Pinsker's (GP) coding to the Gaussian channel and showed that for this channel, interference does not affect the main channel capacity [3]. He chose $U = X + \alpha V$ and maximized the Gelfand and Pinsker's capacity over all quantities of $\alpha$ and proved that for the optimum value of $\alpha$ and for independent Gaussian $X$ and $V$, the capacity of the channel reduces to the primary main channel without side information.

Mitrpant et al., extended the dirty paper channel to the basic Gaussian wiretap channel with side information [4]. They introduced an achievable rate region and an upper bound for this channel. Chen and Vinck investigated Wyner's wiretap channel with side information [5]. They based their results on the previous work on the wiretap channel and the discrete memoryless channel with state information and gave an achievable rate region which is established on the combination of Gelfand–Pinsker coding and the *random coding* used in wiretap channels [6]. They extended their results to the Gaussian wiretap channel with side information using a technique like dirty paper coding. They proved that known channel state at the transmitter can increase the secure rate region in the state-dependent wiretap channel.

The effect of the channel state on the achievable rate in cognitive interference channel was considered in [7]. The coding and the rate region in this paper was based on [8] for the MAC. The achievable rate and the outer bound for the cognitive interference channel in which the cognitive transmitter knows the channel state non-causally was derived in [7]. They proved that this side information make the cognitive encoder to increase the pair rate in a cooperative manner.

The secrecy rate of the cognitive interference channel was studied in [9]. In this work, the cognitive encoder has the primary sender's message and an individual message. At the recievers, the primary receiver acts as a wiretapper for the cognitive one and the aim is to decrease the lekage information to the unintended reciever, i.e., primary reciever. The random coding used and the achievable equivocation rate region was derived. The authors in [10] studied the cognitive interference channel with two confidential messages. In this scenario, each receiver acts as an eavesdropper for the other one. So, the message of the primary and the secondary transmitter must be secure at the unintended destinations. In this model the cognitive encoder, which knows the message of the primary encorder non-causally, acts in a manner that both the messages whould be secure at the unintended recievers.

In this paper we consider the cognitive state-dependent interference channel with a confidential message, in which the channel state models the interfering signals. As we can see in Fig. 1, we have two transmitter-receiver pairs. The secondary transmitter, i.e., the cognitive transmitter knows the message of the primary transmitter and the state of the channel as the side information. Each receiver, decodes its individual message. The message of the cognitive transmitter must be kept confidential in unintended receiver. On the other hand, the primary receiver acts as an eavesdropper for the cognitive encoder.

Not considering the secrecy issue, our model reduces to the model studied in [7]. Furthermore, the cognitive channel without channel state is like the one considered in [9]. We employ the coding scheme used in [5] and establish the equivocation rate region for the cognitive interference channel, which characterizes the tradeoff between the achievable rates and the achievable secrecy at the primary receiver, i.e., the eavesdropper. So, we derive the achievable rate and the outer bound for this channel. This paper is organized as follows. In Section II, the channel model is introduced. In Section III, the main results are presented. The paper is concluded in Section V. The proofs for the theorems are presented in the Appendices.

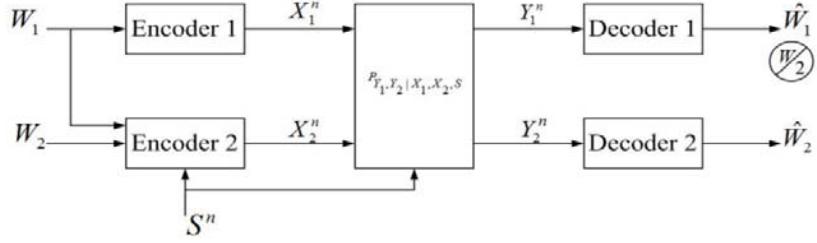

Fig. 1. The cognitive state-dependent interference channel with a confidential message.

## II. CHANNEL MODEL

Consider a memoryless stationary state-dependent interference channel with finite input alphabets $\mathcal{X}_1$ and $\mathcal{X}_2$, finite output alphabets $\mathcal{Y}_1$ and $\mathcal{Y}_2$, the state alphabet $\mathcal{S}$ with distribution $\mathcal{P}_S$ and a conditional probability distribution $P_{Y_1,Y_2|X_1,X_2,S}$. In the sequel, we use $x^n$ and $x_i^n$ to indicate the vectors $(x_1, \cdots, x_n)$ and $(x_i, \cdots, x_n)$, respectively. The $t$-th encoder wishes to transmit the message $W_t$ uniformly distributed on the set $\mathcal{W}_t = \{1, \cdots, M_t\}$, where $t = 1, 2$. The message $W_1$ is known in both encoders but the message $W_2$ is just known at the encoder 2. This encoder assumed as the cognitive transmitter.

The conditional distribution of the channel output n-sequences $Y_1^n, Y_2^n$ given the inputs and the states, $n$-sequences $X_1^n, X_2^n, S^n$ take the product form

$$P_{Y1^n,Y2^n|X1^n,X2^n,S^n}(y_1^n, y_2^n|x_1^n, x_2^n, s^n) = \prod_{i=1}^{n} P_{Y1,Y2|X1,X2,S}(y_{1,i}, y_{2,i}|x_{1,i}, x_{2,i}, s_i) \tag{1}$$

The encoders for the channel are defined by the mappings

$$\varphi_{1,n}: \mathcal{W}_1 \to \mathcal{X}_1$$

$$\varphi_{2,n}: \mathcal{W}_1 \times \mathcal{W}_2 \times S^n \to \mathcal{X}_2^n \tag{2}$$

and the decoders for the channel are defined by the mappings

$$\psi_{1,n}: \mathcal{Y}_1 \to \widehat{\mathcal{W}_1}$$

$$\psi_{2,n}: \mathcal{Y}_2 \to \widehat{\mathcal{W}_2} \tag{3}$$

We denote the error probability $P_e^{(n)} = \max(P_{e,1}^{(n)}, P_{e,2}^{(n)})$ in which

$$P_{e,t} = \sum_{w_1,w_2} \frac{1}{M_1 M_2} P[\psi_{t,n} \neq W_t | W_t \text{ sent}] \tag{4}$$

*Definition 1:* We define $R_i = \left\lfloor \frac{1}{n} \log M_i \right\rfloor$ for $t = 1, 2$.

*Definition 2:* The rate-triple $(R_1, R_2, R_{e2})$ is achievable if for any $\epsilon_n > 0$ there exists a $(M_1, M_2, n, P_n)$ code such that $P_e \leq \epsilon$.

The secrecy level of the secondary encoder's message at the primary receiver is measured by normalized equivocation

$$R_{e2}^{(n)} = \frac{1}{n} H(W_2|Y_1), \quad (5)$$

and

$$0 \leq R_{e2} \leq \liminf_{n \to \infty} R_{e2}^{(n)}, \quad (6)$$

for $t = 1, 2$.

*Definition 3:* The capacity region is the closure of the set of all achievable rate-triples.

III. MAIN RESULTS: INNER AND OUTER BOUNDS

As we explained in the previous section, in our scenario the cognitive encoder has non-causal access to the message of the primary sender. In addition, the state of the channel is assumed to be known at the cognitive transmitter. The following results give the achievable region for the finite alphabet cognitive interference channel with CSIT.

*Theorem 1:* (achievable rate) The closure of the convex hull of the set of rate-triples ($R_1$, $R_2$, $R_{e2}$) satisfying

$$R_1 \leq I(U, X_1; Y_1) - I(U, X_1; S) \quad (7)$$
$$R_2 \leq I(V; Y_2) - I(V; U, X_1, S) \quad (8)$$
$$R_{e2} \leq I(V; Y_2) - \max\{I(V; U, X_1, S), I(V; U, X_1, Y_1)\} \quad (9)$$

for input distribution factors as

$$P_{S,U,V,X_1,X_2,Y_1,Y_2} = P_S P_{X_1} P_{U,V,X_2|S,X_1} P_{Y_1,Y_2|S,X_1,X_2} \quad (10)$$

is achievable for the finite alphabet cognitive interference channel with CSIT.

*Proof:* The details on the proof and the computation of the equivocation rate are relegated to the Appendix A. As a hint, the cognitive encoder uses the mutual information between its message $X_2$ and the state of the channel and the output signal in the receiver 1, to randomize its message. Using the random coding, message of encoder 2 remains confidential at the receiver 1 and by Gelfand and Pinsker coding scheme the effect of the channel state is canceled.

*Corollary 1:* We note that this result without secrecy issue reduces to the result of [6, Th. 1] for the cognitive state-dependent interference channel.

*Corollary 2:* The equivocation-rates (7)-(9) without channel state, i.e., in the case that $S = \emptyset$, is reduced to the result derived in (6) of [9].

*Corollary 3:* In the special case, when we have $I(V; U, X_1, S) > I(V; U, X_1, Y_1)$, the achievable rate region (7)-(9) reduces to the achievable rate region of the state-dependent cognitive interference channel in [7]. In this case, the coding scheme used in [7], achieves the secure rate for the cognitive transmitter.

*Corollary 4:* In a special case, when $I(V; U, X_1, S) < I(V; U, X_1, Y_1)$, the achievable rate region (7)-(9) reduces to the following remark.

*Remark 1:* The closure of the convex hull of the set of rate-pairs $(R_1, R_2)$ satisfying

$$R_1 \leq I(U, X_1; Y_1) - I(U, X_1; S) \tag{11}$$

$$R_2 \leq I(V; Y_2) - I(V; U, X_1, Y_1) \tag{12}$$

for input distribution factors as (10), is an achievable secure rate region for the finite alphabet cognitive interference channel with CSIT in special case, when $I(V; U, X_1, S) < I(V; U, X_1, Y_1)$.

*Theorem 2:* (outer bound) The set of achievable rate-triples of the cognitive interference channel with CSIT is contained in the closure of the set rate-triple $(R_1, R_2, R_{e2})$ that satisfy

$$R_1 \leq I(U, X_1; Y_1) - I(U, X_1; S), \tag{13}$$

$$R_2 \leq I(X_2; Y_2 | X_1, S), \tag{14}$$

$$R_{e2} \leq \min \left\{ \begin{array}{c} I(X_2; Y_2 | X_1, S) - I(X_2; Y_1 | X_1, S), \\ I(X_2; Y_2 | U, X_1, S) - I(X_2; Y_1 | U, X_1, S) \end{array} \right\}, \tag{15}$$

for input distribution factors as (10).

*Proof:* We relegate the details on the proof of the above theorem to the Appendix B, but we should note that the equations (13) and (14) are equal to the result in [7, Theorem 2], i.e., the outer bound in Theorem 2 is reduced to the result in [7] without secrecy issue. The equation (15) is derived with the technique presented in [10].

## IV. CONCLUSIONS

In this paper the secrecy problem of the state-dependent cognitive interference channel with non-causally CSIT was considered. The achievable rate region and the outer bound on the rate region was derived. The results are compared with the previous works in the special cases.

## APPENDIX A: PROOF OF THE THEOREM 1

In this Appendix, we present the proof of the Theorem 1. First of all, we introduce the coding scheme. In this section, we denote the messages by $W_1 \in \mathcal{W}_1 = \{1,2,\ldots,2^{nR_1}\}$ and $W_2 \in \mathcal{W}_2 = \{1,2,\ldots,2^{nR_2}\}$, respectively. The notation $T_\epsilon^n(P_{UX_1X_2Y_1Y_2})$ is used to indicate the strong typical set based on the distribution $P_{UX_1X_2Y_1Y_2}$.

### A. Code Generation:

The code generation scheme is as following

1. Generate $2^{nR_1}$ codeword $u_i^n$, $i = \{1,2,\ldots,2^{nR_1}\}$, each uniformly drawn from the set $T_\epsilon^n(P_U)$.

2. For each $u_i^n$, generate $2^{nR_2}$ sequences $v_{i,k}^n$, $k = \{1,2,\ldots,2^{nR_2}\}$, each uniformly drawn from the set $T_\epsilon^n(P_{V|U})$.

3. For using the random coding scheme to keep the cognitive message be secret at the unintended receiver, distribute the $v^n$ sequences randomly to $2^{nR}$ bin such that each bin contains $2^{n[\max\{I(X_2;Y_1,U,X_1),\ I(X_2;S,U,X_1)\}]}$ sequences and $R = R_2 - \max\{I(X_2;Y_1,U,X_1),\ I(X_2;S,U,X_1)\}$. Now, we index each bin by $j \in \{1,2,\cdots,2^{nR}\}$. Next, partition $2^{n[\max\{I(X_2;Y_1,U,X_1),\ I(X_2;S,U,X_1)\}]}$ sequences in every bin into $2^{n[\max\{I(X_2;Y_1,U,X_1),\ I(X_2;S,U,X_1)\}-I(X_2;Y_1,U,X_1)]}$ subbin each subbin contains $2^B = 2^{n[I(X_2;Y_1,U,X_1)]}$ sequences. Index each subbin by $a \in \{1,2,\ldots,2^{n[\max\{I(X_2;Y_1,U,X_1),I(X_2;S,U,X_1)\}-I(X_2;Y_1,U,X_1)]}\}$ and let $A$ be the random variable to represent the index of the subbin. This step of code generation is illustrated in Fig 2.

### B. Encoding:

Now, we define $\mathcal{A} = \{1,2,\ldots,A\}$, $\mathcal{B} = \{1,2,\ldots,B\}$ where $A$ and $B$ were defined before. So, we let $W_2 = \mathcal{A} \times \mathcal{C}$ where $\mathcal{C} = \{1,2,\ldots,B\}$. We define mapping $\mathcal{B} \to \mathcal{C}$ to be partitioning $\mathcal{B}$ into $\mathcal{C}$ subsets with nearly equal size, where "nearly equal size" means

$$\|g^{-1}(j_1)\| \leq 2\|g^{-1}(j_2)\|, \forall j_1, j_2 \in \mathcal{C} \quad (16)$$

Encoder 1 chooses $u_i^n$ according to its message index $i$. Encoder 2, tries to find the $k$ such that $(u_i^n, v_{i,k}^n) \in T_\epsilon^n$. This index, contains the individual message of the cognitive encoder. Then, it finds $j$, $a$ and $b$ such that $(v_{i,k,j,a,b}^n, u_i^n, s^n) \in T_\epsilon^n$. So, it lets $W_2 = (a,c) \to (a,b)$ with $b$ randomly, uniformly from the set $g^{-1}(j) \subset \mathcal{B}$. Encoder 1 and 2 send $x_{1,i}^n$ and $x_{2,i,k,j,a,b}^n$ according to $p(x_{1,i}^n|u_i^n) = \prod_{i=1}^n p(x_{1,i}|u_i)$ and $p(x_{2,i,k,j,a,b}^n|v_{i,k,j,a,b}^n, u_i^n, s^n) = \prod_{i=1}^n p(x_{2,i,k,j,a,b}|v_{i,k,j,a,b}, u_i, s_i)$, respectively.

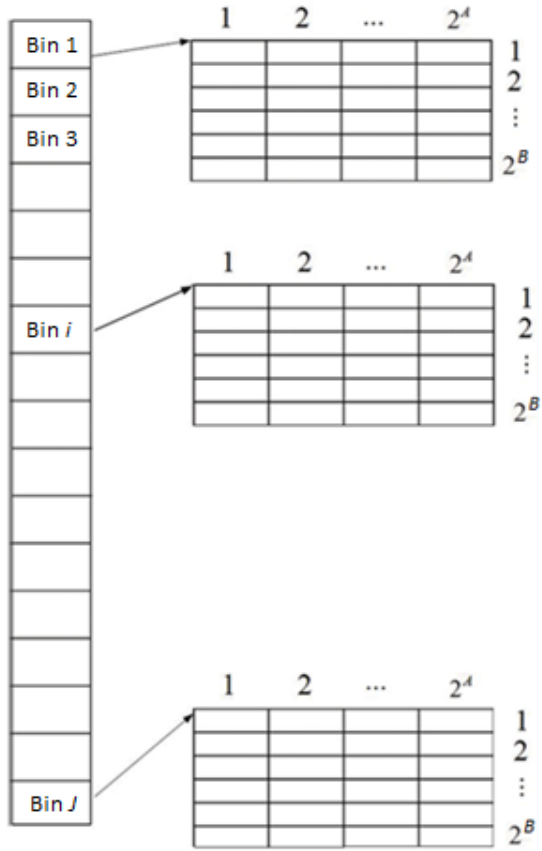

Fig. 2. Code generation scheme for the cognitive interference channel with CSIT.

*C. Decoding:*

Receiver 1 tries to find $\hat{i}$ such that $(x_{1,\hat{i}}^n, y_1^n) \in T_\epsilon^n$. If no such $\hat{i}$ is found, the error is declared.

Receiver 1, if given the indices $i, k, j, a, b$, declares that the index of $x_{2,i,k,j,a,b}^n$ is $\hat{b}$, if it is a unique index such that $(x_{1,i}, x_{2,i,k,j,a,\hat{b}}, y_1^n) \in T_\epsilon^n$. If no such $\hat{b}$ is found, an error is declared.

Receiver 2, tries to find the $\hat{i}, \hat{k}, \hat{j}, \hat{a}, \hat{b}$ such that $(x_{2,\hat{i},\hat{k},\hat{j},\hat{a},\hat{b}}, u_i^n, y_2^n) \in T_\epsilon^n$. If there are not such indices, the error is indicated.

*D. Error Computation:*

We can compute the average error by the standard techniques as in [11], where the average is taken over the random codebook ensembles. So, we can show that the average error is less than $\lambda$ for sufficiently large codeword length $n$.

*E. Equivocation Computation:*

In this section, we compute the equivocation of the $W_2$ at the receiver 1, in following:

$$\begin{aligned}
H(W_2|Y_1^n) &= H(W_2, Y_1^n) - H(Y_1^n) = H(W_2, Y_1^n, A, W_1) - H(Y_1^n) - H(A, W_1|W_2, Y_1^n)\\
&= H(W_2, A, W_1, Y_1^n, X_2^n) - H(X_2^n|W_2, A, W_1, Y_1^n) - H(Y_1^n) - H(A, W_1|W_2, Y_1^n)\\
&= H(W_2, A, W_1|Y_1^n, X_2^n) + H(Y_1^n, X_2^n) - H(X_2^n|W_2, A, W_1, Y_1^n) - H(Y_1^n)\\
&\quad - H(A, W_1|W_2, Y_1^n) \overset{(a)}{\geq} H(X_2^n|Y_1^n) - H(X_2^n|W_2, A, W_1, Y_1^n)\\
&\quad - H(A, W_1|W_2, Y_1^n) \overset{(b)}{\geq} H(X_2^n|Y_1^n) - H(X_2^n|W_2, A, W_1, Y_1^n) - \log|\mathcal{A}|\\
&\quad - H(X_2^n|Y_2^n) \overset{(c)}{\geq} n[I(X_2; Y_2) - I(X_2; Y_1)] - H(X_2^n|W_2, A, W_1, Y_1^n)\\
&\quad - n[\max\{I(X_2; Y_1, U, X_1), I(X_2; S, U, X_1)\}\\
&\quad - I(X_2; S, U, X_1)] \overset{(d)}{\geq} n[I(X_2; Y_2) - \max\{I(X_2; Y_1, U, X_1), I(X_2; S, U, X_1)\}]\\
&\quad - H(X_2^n|W_2, A, W_1, Y_1^n)
\end{aligned} \quad (17)$$

where

$(a)$ follows from the fact that we have $H(W_2, A, W_1, Y_1^n, X_2^n) \geq 0$;

$(b)$ follows from the fact that $H(A, W_1|W_2, Y_1^n) = H(A|W_2, Y_1^n) + H(W_1|A, W_2, Y_1^n) \leq H(A) \leq \log|\mathcal{A}|$, because of the fact that $H(W_1|A, W_2, Y_1^n) = 0$;

$(c)$ follows from the fact that for the $n \to \infty$, we have $I(X_2^n; Y_1^n) = n\, I(X_2; Y_1)$, $I(X_2^n; Y_2^n) = n\, I(X_2; Y_2)$ and $\log|\mathcal{A}| = n[\max\{I(X_2; Y_1, U, X_1), I(X_2; S, U, X_1)\} - I(X_2; Y_1, U, X_1)]$;

$(d)$ is because of the fact that $I(X_2; Y_1, U, X_1) - I(X_2; Y_1) = I(X_2; U, X_1|Y_1) \geq 0$.

Now, for the last term in (17), we define

$$\rho(w_1, w_2, a, y_1^n) = \begin{cases} x_{2,i,k,j,a,b} & \text{if we have } (x_{1,i}, x_{2,i,k,j,a,b}, y_1^n) \in T_\epsilon^n\\ x_{2,1,1,1,1,1} & \text{otherwise} \end{cases} \quad (18)$$

Then

$$Pr\{X_2^n \neq \rho(W_1, W_2, A, Y_1^n)\} = \sum_{w_1, w_2, a, b} [P_{w_1, w_2, a, b} \cdot \Pr\{x_{2,i,k,j,a,b}\\ \neq \rho(w_1, w_2, a, y_1^n)|w_1, w_2, a, b\}] = \lambda < \eta. \quad (19)$$

Therefore, by Fano's inequality, we obtain

$$\frac{1}{n}H(X_2^n|W_2,A,W_1,Y_1^n) \leq \frac{1}{n}(1+\lambda\log(2^{n(R_1+R_2)})) \leq \epsilon \tag{20}$$

where $\epsilon$ is small for sufficiently large $n$.

By combining (6), (17) and (20), the achievable rate region in Theorem 2 is derived. ∎

APPENDIX B: PROOF OF THE THEOREM 2

Proof: As we mentioned later, the equations (13) and (14) in Theorem 2 are similar to the results derived by [7]. In this section we prove (15) in Theorem 2. First we define the following notations

$$y^i \triangleq (y_1, y_2, \ldots, y_i) \tag{21}$$

$$\tilde{y}^i \triangleq (y_i, y_{i+1}, \ldots, y_n) \tag{22}$$

Now, we consider the equivocation rate bound applying the techniques in [12]

$$R_{e2} \leq H(w_2|y_1^n) = H(w_2) - I(w_2; y_1^n)$$
$$= I(w_2; y_2^n) - I(w_2; y_1^n) + H(w_2|y_2^n) \leq^{(e)} I(w_2; y_2^n) - I(w_2; y_1^n) \tag{23}$$
$$+ 2n\epsilon_n,$$

$$R_{e2} \leq H(w_2|y_1^n) = H(w_2|w_1, y_1^n) - I(w_2; w_1|y_1^n)$$
$$\leq H(w_2|w_1) - I(w_2; y_1^n|w_1) + H(w_2|y_2^n)$$
$$= I(w_2; y_2^n|w_1) - I(w_2; y_1^n|w_1) + H(w_2|y_2^n, w_1) \tag{24}$$
$$+ H(y_1^n|w_1) \leq^{(f)} I(w_2; y_2^n|w_1) - I(w_2; y_1^n|w_1) + 2n\epsilon_n$$

where $(e)$ and $(f)$ are because of Fano's inequality and $\epsilon_n$ is negligible

for enough large $n$. So, we have

$$I(w_2; y_1^n) = \sum_{i=1}^n I(w_2; y_{1i}|y_2^{i-1}, \tilde{y}_2^{i+1}) + \sum_{i=1}^n I(y_2^{i-1}; y_{1i}|\tilde{y}_2^{i+1}) - \sum_{i=1}^n I(y_2^{i-1}; y_{1i}|\tilde{y}_2^{i+1}, w_1, w_2). \tag{25}$$

$$I(w_2; y_1^n|w_2) = \sum_{i=1}^n I(w_2; y_{1i}|y_2^{i-1}, \tilde{y}_1^{i+1}, w_1) + \sum_{i=1}^n I(y_2^{i-1}; y_{1i}|\tilde{y}_1^{i+1}, w_1)$$
$$- \sum_{i=1}^n I(y_2^{i-1}; y_{1i}|\tilde{y}_1^{i+1}, w_2, w_1) \tag{26}$$

Therefore we have

$$R_{e2} \leq \sum_{i=1}^n I(w_2; y_{2i}|y_2^{i-1}, \tilde{y}_1^{i+1}) - \sum_{i=1}^n I(w_2; y_{1i}|y_2^{i-1}, \tilde{y}_1^{i+1}) + 2n\epsilon_n \tag{27}$$

$$R_{e2} \leq \sum_{i=1}^n I(w_2; y_{2i}|y_2^{i-1}, \tilde{y}_1^{i+1}, w_1) - \sum_{i=1}^n I(w_2; y_{1i}|y_2^{i-1}, \tilde{y}_1^{i+1}, w_1) + 2n\epsilon_n \tag{28}$$

Let us define random variable $T$, independent of $w_1, w_2, x_1^n, x_2^n, y_1^n, y_2^n$

$$V \triangleq y_2^{T-1} \tilde{y}_1^{T+1} T, X_2 \triangleq x_{2T}, Y_2 \triangleq y_{2T}, Y_1 \triangleq y_{1T}, U_2 \triangleq w_2, U_1 \triangleq w_1 \tag{29}$$

Substituting these variables we obtain (20). ∎

REFERENCES

[1] C. Shannon, "Channels with side information at the transmitter," *J. Res. Devel.*, vol. 2, pp. 289-293, 1958.

[2] S. I. Ge'lfand and M. S. Pinsker, "Coding for channel with random parameters," *Probl. Inf. Theory*, vol. 9, no. 1, pp. 19–31, 1980.

[3] M. H. M. Costa, "Writing on dirty paper*,*" *IEEE Trans. Inf. Theory*, vol. IT-29, no. 3, pp. 439-441, May 1983.

[4] C. Mitrpant, H. Vinck, and Y. Luo, "An achievable region for the Gaussian wiretap channel with side information," *IEEE Trans. Inf. Theory*, vol. 52, no. 5, pp. 2181–2190, 2006.

[5] Y. Chen and H. Vinck, "Wiretap channel with side information," *IEEE Trans. Inf. Theory*, vol. 54, no. 1, pp. 395–402, 2008.

[6] A. Wyner, "The wire-tap channel," *Bell. Syst. Tech. J.*, vol. 54, no. 8, pp. 1355–1387, Jan. 1975.

[7] A. Somekh-Baruch, S. Shamai (Shitz), and Verdu Sergio, "Cognitive interference channels with state information," in *Int. Symp. Inf. Theory (ISIT)*, Toronto, Canada, July 6–11, 2008, pp. 1353–1357.

[8] A. Somekh-Baruch, S. Shamai (Shitz), and Verdu Sergio, "Cooperative multiple-access encoding with states available at one transmitter," *IEEE Trans. Inform. Theory*, vol. 54, no. 10, pp. 4448–4469, Oct. 2008.

[9] Y. Liang, A. Somekh-Baruch, H. Vincent Poor, Shlomo Shamai (Shitz), and Sergio Verdo, "Capacity of cognitive interference channels with and without secrecy," *IEEE Trans. Inf. Theory*, vol. 55, no. 2, pp. 604–618, Feb 2009.

[10] H. G. Bafghi, S. Salimiy, B. Seyfe, and M. R. Aref, "Cognitive interference channel with two confidential messages," in *Int. Symp. Inf. Theory and Applic. (ISITA)*, 2010, pp. 952–956.

[11] T. M. Cover and J. A. Thomas, *Elements of Information Theory*, J. Wiley and Sons, Inc., 2006.

[12] I. Csiszar and J. Korner, "Broadcast channels with confidential messages," *IEEE Trans. on Inf. Theory*, vol. 24, no. 3, pp. 339–348, May 1978.